\documentclass[11pt]{article}
\usepackage[textwidth=15.2cm,textheight=22cm]{geometry}
\usepackage{amsmath,amssymb}
\usepackage{latexsym}
\usepackage{multicol}
\usepackage{graphicx}
\usepackage{bm}
\allowdisplaybreaks[1]

\newcommand{\be}{\begin{equation}}
\newcommand{\ee}{\end{equation}}
\newcommand{\ba}{\begin{eqnarray}}
\newcommand{\ea}{\end{eqnarray}}
\newcommand{\bdm}{\begin{displaymath}}
\newcommand{\edm}{\end{displaymath}}

\newcommand\fr[1]{\frac{1}{#1}}

\newcommand{\rom}[1]{\uppercase\expandafter{\romannumeral #1\relax}}

\def\ba{\bar A}

\def\beq{\begin{equation}}
\def\eeq{\end{equation}}

\newcommand{\nn}{\nonumber}

\newcommand{\ndt}{\noindent}

\newcommand{\delp}{{\partial^+}}

\def\bea{\begin{eqnarray}}
\def\eea{\end{eqnarray}}
\def\beas{\begin{eqnarray*}}
\def\eeas{\end{eqnarray*}}
\def\sla{\raise.15ex\hbox{$/$}\kern-.57em}

\def\parp{\partial_+}

\def\spa#1.#2{\left\langle#1\,#2\right\rangle}
\def\spb#1.#2{\left[#1\,#2\right]}

\begin{document}

\begin{titlepage}
\begin{flushright}    
{\small $\,$}
\end{flushright}
\vskip 1cm
\centerline{\Large{\bf{Yang-Mills theories}}}
\vskip 0.5cm
\centerline{\Large{\bf{and quadratic forms}}}
\vskip 1.5cm
\centerline{Sudarshan Ananth$^*$, Lars Brink$^\dagger$ and Mahendra Mali$^*$}
\vskip .5cm
\centerline{$*$\it {Indian Institute of Science Education and Research}}
\centerline{\it {Pune 411008, India}}
\vskip .5cm
\centerline{$\dagger$\it {Department of Fundamental Physics}}
\centerline {\it {Chalmers University of Technology}}
\centerline{\it {S-412 96 G\"oteborg, Sweden}}
\vskip 1.5cm
\centerline{\bf {Abstract}}
\vskip .5cm
We show that the Hamiltonian of $(\mathcal N=1, d=10)$ super Yang-Mills can be expressed as a quadratic form in a very similar manner to that of the $(\mathcal N=4, d=4)$ theory. We find a similar quadratic form structure for pure Yang-Mills theory but this feature, in the non-supersymmetric case, seems to be unique to four dimensions. We discuss some consequences of this feature.
\vfill
\end{titlepage}

\section{Introduction}

\ndt Symmetries play a crucial role in quantum field theories. They determine whether a quantum field theory is consistent or not and further determine the physics of the theory. In this respect, supersymmetry plays an even more penetrating role than ordinary symmetries. It directly affects the perturbative properties of the theory since bosons and fermions behave differently in loop diagrams. The $SO(8)$ triality of $\mathcal N=1$ super Yang-Mills theory in ten dimensions, for example, results in the ultra-violet finiteness of its four-dimensional avatar, $\mathcal N=4$ Yang-Mills theory leaving it conformally invariant in the quantum regime. The lack of such manifest beauty in $SO(9)$ appears to doom $\mathcal N=8$ supergravity to a divergent end. However, the light-cone descriptions of the $\mathcal N=4$ and $\mathcal N=8$ theories~\cite{Brink:1982pd, Bengtsson:1983pg} are so similar that they seem to hint at a deep link between them. In addition, there are the KLT-relations~\cite{KLT} which suggest that the finiteness properties of the $\mathcal N=4$ model could possibly extend to the $\mathcal N=8$ model. Even if this turns out not to be the case, the perturbative properties of these theories are so remarkable that one might ask if there exist additional algebraic properties that affect the quantum behavior of these theories.
\vskip 0.3cm

\ndt The question then is how much of the unique quantum behavior of maximally supersymmetric theories is due entirely to the maximal supersymmetry. Are there more to these theories than meets the eye? In this paper, we describe how the Hamiltonians of both the maximally supersymmetric Yang-Mills theory in four dimensions and its higher-dimensional parent may be expressed as quadratic forms and we argue that it is only the maximally supersymmetric theories, among the supersymmetric ones, that exhibit this property.  We then turn to pure non-supersymmetric Yang-Mills theory where we illustrate why this feature, of quadratic forms, occurs only in a helicity basis.
\vskip 0.3cm
\ndt Unlike the $\mathcal N=4$ theory which possesses conformal invariance, $\mathcal N=8$ supergravity has an on-shell non-linear $E_{7(7)}$ symmetry. A key question in this context, is whether this exceptional symmetry is associated specifically with dimensional reduction or whether it is simply a reflection of its higher dimensional parent. This question is one of the motivations for our present line of work and this paper, an intial step in this study. The other motivation is what was outlined above, to identify new algebraic properties of maximally supersymmetric quantum field theories that can further constrain these theories perturbatively and non-perturbatively.
\vskip 0.3cm
\ndt Like the $\mathcal N=4$ theory, the $\mathcal N=8$ Hamiltonian, in light-cone gauge, may be written as a quadratic form, involving the dynamical supersymmetry generator~\cite{ABHS}. If we could successfully `lift' or `oxidize' this quadratic form structure to eleven dimensions, then the eleven-dimensional theory would also exhibit $E_{7(7)}$ invariance assuming that the `lifting' process commutes with the exceptional generators. This is work in progress.
\vskip 0.3cm
\ndt In this paper, we explain the idea of the quadratic form structure and our methods, in the context of the Yang-Mills system. We start with the supersymmetric theories and then move to pure Yang-Mills. In these considerations we will also highlight the importance of residual gauge invariance within light-cone gauge.
 
\section{Super Yang-Mills theory}

\subsection{($\mathcal N=4,d=4$) Yang-Mills theory in light-cone superspace}

\ndt This section serves as a brief review of the results in~\cite{Brink:1982pd,ABR1} relevant to this paper. The $\mathcal N=4$ theory involves one complex bosonic field (the gauge field), four complex Grassmann fields and six scalar fields. The form of the theory that we will use here can be obtained in two ways. The first method is to choose light-cone gauge, $A^+=\frac{1}{\sqrt 2}(A^0+A^3)=0$ and then solve for the unphysical field $A^-=\frac{1}{\sqrt 2}(A^0-A^3)$. The other approach is to start with the superfield, introduced below, and span the super Poincar\'e algebra on it. Since the representation is non-linear, certain generators will be non-linear, including the generator $P^-=\frac{1}{\sqrt 2}(P^0-P^3)$ which is the full Hamiltonian. With the metric $(-,+,+,\dots,+)$, the light-cone coordinates and derivatives are $x^\pm\;;\;\partial_\pm$ and 
\bea
x=\frac{1}{\sqrt 2}\,(\,{x_1}\,+\,i\,{x_2}\,)\ ;\qquad  {\bar\partial} =\frac{1}{\sqrt 2}\,(\,{\partial_1}\,-\,i\,{\partial_2}\,)\ .
\eea
With the introduction of anticommuting Grassmann variables $\theta^m$ and $\bar\theta_m$ ($m,n,p,q,\dots =1,2,3,4$, denote $SU(4)$ spinor indices), {\it all} the physical degrees of freedom  can be captured  in one superfield 
\bea
\phi\,(y)&=&\frac{1}{ \partial^+}\,A\,(y)\,+\,\frac{i}{\sqrt 2}\,{\theta_{}^m}\,{\theta_{}^n}\,{\bar C^{}_{mn}}\,(y)\,+\,\frac{1}{12}\,{\theta_{}^m}\,{\theta_{}^n}\,{\theta_{}^p}\,{\theta_{}^q}\,{\epsilon_{mnpq}}\,{\partial^+}\,{\bar A}\,(y)\cr
& &~~~ +~\frac{i}{\partial^+}\,\theta^m_{}\,\bar\chi^{}_m(y)+\frac{\sqrt 2}{6}\theta^m_{}\,\theta^n_{}\,\theta^p_{}\,\epsilon^{}_{mnpq}\,\chi^q_{}(y) \ .
\eea
where $A$ and $\bar A$ represent the gauge field and the six scalar fields, written as antisymmetric $SU(4)$ bi-spinors, satisfy 
\bea
\label{dual}
{{\bar C}^{}_{mn}}~=~\,\frac{1}{2}\,{\epsilon^{}_{mnpq}}\,{C_{}^{pq}} \ .
\eea
The fermion fields are denoted by $\chi^m$ and $\bar\chi_m$. All fields carry adjoint indices (not shown here), and are local in the coordinates  
\bea
y~=~\,(\,x,\,{\bar x},\,{x^+},\,y^-_{}\equiv {x^-}-\,\frac{i}{\sqrt 2}\,{\theta_{}^m}\,{{\bar \theta}^{}_m}\,)\ .
\eea
\ndt The chiral derivatives
\bea
{d^{\,m}}=-{\partial^m}\,-\,\frac{i}{\sqrt 2}\,{\theta^m}\,{\partial^+}\ ;\qquad{{\bar d}_{\,n}}=\;\;\;{{\bar \partial}_n}\,+\,\frac{i}{\sqrt 2}\,{{\bar \theta}_n}\,{\partial^+}\ ,
\eea
satisfy
\be
\{\,{d^m}\,,\,{{\bar d}_n}\,\}\,=\,-i\,{\sqrt 2}\,{{\delta^m}_n}\,{\parp}\ .
\ee
The superfield $\phi$ and its complex conjugate $\bar\phi$ satisfy
\be
{d^{\,m}}\,\phi\,=\,0\ ;\qquad {\bar d_{\,m}}\,\bar\phi\,=\,0\ ,
\ee
and the ``inside-out" constraints
\bea
\label{ioc}
\bar d_m^{}\,\bar d_n^{}\,\phi~=~\frac{1}{ 2}\,\epsilon_{mnpq}^{}\,d^p_{}\,d^q_{}\,\bar\phi\ , \nn \\
 d^m_{}\, d^n_{}\,\bar\phi~=~\frac{1}{ 2}\,\epsilon^{mnpq}_{}\,\bar d_p^{}\,\bar d_q^{}\,\phi\ .
\eea
In terms of this superfield, the $(\mathcal N=4, d=4)$ action is
\be
\label{fourdd}
\int d^4x\int d^4\theta\,d^4 \bar \theta\;{\cal L}\ ,
\ee
where
\bea \label{4Lag}
{\cal L}&=&-\bar\phi\,\frac{\Box}{\partial^{+2}}\,\phi
~+\frac{4g}{3}\,f^{abc}_{}\,\Big(\frac{1}{\partial^+_{}}\,\bar\phi^a_{}\,\phi^b_{}\,\bar\partial\,\phi^c_{}+{\rm complex~conjugate}\Big)\cr
&&-g^2f^{abc}_{}\,f^{ade}_{}\Big(\,\frac{1}{\partial^+_{}}(\phi^b\,\partial^+\phi^c)\frac{1}{\partial^+_{}}\,(\bar \phi^d_{}\,\partial^+_{}\,\bar\phi^e)+\frac{1}{2}\,\phi^b_{}\bar\phi^c\,\phi^d_{}\,\bar\phi^e\Big)\ .
\eea
Grassmann integration  is normalized so that $\int d^4\theta\,\theta^1\theta^2\theta^3\theta^4=1$, and $f^{abc}$ are the structure functions of the  Lie algebra.
\vskip 0.3cm
\ndt The supersymmetry generators are of two varieties: kinematical and dynamical. The kinematical (spectrum generating) supersymmetries
\be
q^{\,m}_{\,+}=-{\partial^m}\,+\,\frac{i}{\sqrt 2}\,{\theta^m}\,{\partial^+}\ ;\qquad{{\bar q}_{\,+\,n}}=\;\;\;{{\bar \partial}_n}\,-\,\frac{i}{\sqrt 2}\,{{\bar \theta}_n}\,{\partial^+}\ ,
 \ee
satisfy
\be
\{\,q^{\,m}_{\,+}\,,\,{{\bar q}_{\,+\,n}}\,\}\,=\,i\,{\sqrt 2}\,{{\delta^m}_n}\,{\delp}\ ,
\ee
while the dynamical supersymmetries are obtained by boosting the kinematical ones
\be
{q}^m_{\,-}~\equiv~i\,[\,\bar j^-_{}\,,\,q^{\,m}_{\,+}\,]~=~\frac{\partial}{\partial^+_{}}\, q^{\,m}_{\,+}\ ,\qquad 
{\bar{q}}_{\,-\,m}^{}~\equiv~i\,[\, j^-_{}\,,\,\bar q_{\,+\,m}^{}\,]~=~\frac{\bar\partial}{\partial^+_{}}\, \bar q_{\,+\,m}^{}\ ,
\ee
and satisfy  the  free $N=4$ supersymmetry algebra
\be
\{\, {q}^m_{\,-}\,,\,{\bar{q}}_{\,-\,n}^{}\,\}~=~i\,\sqrt{2}\,\delta^{\,m}_{~~n}\,\frac{\partial\bar\partial}{\partial^+_{}}\ .
\ee
In the free (linear) theory the generators act directly on the superfield.
\be
\delta_O = O \phi.
\ee
For the dynamical (non-linear) generators, we have to find non-linear terms such that the algebra closes. For the dynamical supersymmetry, the result is 
\be \label{eq:nlinear}
\delta^{}_{\bar q_-{}_m}\,\phi^a_{}~=~\frac{1}{\partial^+}\,\left\{\,(\bar\partial\,\delta^{ab}_{}-gf^{abc}_{}\partial^+_{}\,\phi^c\,)\,\delta^{}_{\bar q_+{}_m}\,\phi^c\,\right\}\ .\ee
\vskip 0.1cm
\ndt Consider now the Hamiltonian that we get from \eqref{fourdd}
\bea\label{hamil}
H\!\!\!&=& \!\!\!\int{d^3}x\,{d^4}\theta\,{d^4}{\bar \theta}\,\left\{
\bar\phi^a_{}\,\frac{2\partial\bar\partial}{\partial^{+2}}\,\phi^a_{}
~-\frac{4}{3}\,g\,f^{abc}_{}\,\Big(\frac{1}{\partial^+_{}}\,\bar\phi^a_{}\,\phi^b_{}\,\bar\partial\,\phi^c_{}+\frac{1}{\partial^+_{}}\,\phi^a_{}\,\bar\phi^b_{}\,\partial\,\bar\phi^c_{}\Big)
\right.\cr 
&&\left.+\,g^2f^{abc}_{}\,f^{ade}_{}\Big( \frac{1}{\partial^+}
(\phi^b\,\partial^+\phi^c)\frac{1}{\partial^+_{}}\,
(\bar\phi^d_{}\,\partial^+_{}\,\bar\phi^e)
+\frac{1}{2}\,\phi^b_{}\bar\phi^c\,\phi^d_{}\,\bar\phi^e\Big)\right\}\ ,
\eea
Using the form in \eqref{eq:nlinear}, it can be written as~\cite{ABKR}
\be \label{4hamil2}
H\,=\,\frac{i}{\sqrt 2}\int{d^3}x\,{d^4}\theta\,{d^4}{\bar \theta}\, \delta_{q^m}\bar \phi^a \,\frac{1}{\partial^+} \,\delta_{\bar q_m} \phi^a.
\ee
This is the remarkable quadratic form alluded to above. The key ingredient in proving this is the use of the "inside-out constraint"~\eqref{ioc}. This point is important since it implies that other supersymmetric Yang-Mills theories cannot be expressed as simple quadratic forms since those theories have no such constraint on the superfield.

\subsection{Ten dimensions}
\ndt The key result in~\cite{ABR1} is that the action for the ten-dimensional $\mathcal N=1$ super Yang-Mills theory can be obtained by simply `oxidizing' (\ref {fourdd}). This is achieved in three steps. First, the introduction of the six extra coordinates and their derivatives, again as antisymmetric bi-spinors
\be
{x^{m\,4}}\,=\,\frac{1}{\sqrt 2}\,(\,{x_{m\,+\,3}}\,+\,i\,{x_{m\,+\,6}}\,)\ ,\qquad 
{\partial^{m\,4}}\,=\,\frac{1}{\sqrt 2}\,(\,{\partial_{m\,+\,3}}\,+\,i\,{\partial_{m\,+\,6}}\,)\ ,
\ee
for $m\ne 4$ and their complex conjugates 
\be{{\bar x}_{pq}}\,=\,\frac{1}{2}\,{\epsilon_{pqmn}}\,{x^{mn}}\ ;\qquad{{\bar \partial}_{pq}}\,=\,\frac{1}{2}\,{\epsilon_{pqmn}}\,{\partial^{mn}}\ .
\ee
Second, making all fields dependent on the extra coordinates. The kinematical supersymmetries  $q^n_+$ and $\bar q^{}_{+n}$, are now assembled into one $SO(8)$ spinor. The dynamical supersymmetries are obtained by boosting 
\be
i\,[\,{\bar J}^-\,,\,q_+^m\,]~\equiv~{\cal Q}^m_{}\ ,\qquad 
i\,[\,J^-\,,\,\bar q_{+\,m}\,]~\equiv~\bar{\cal Q}_m^{}\ ,
\ee
where the linear part of the dynamical boosts are
\bea
J^-_{}&=&i\,x\,\frac{\partial\bar\partial\,+\,{\frac {1}{4}}\,{{\bar \partial}_{pq}}\,{\partial^{pq}}}{\partial^+_{}} ~-~i\,x^-_{}\,\partial+i\,{\frac { \partial}{\parp}}\,\Big\{\,{ \theta}^m\,{\bar\partial_m}~+~\frac{i}{4\sqrt{2}\,\parp}\,(d^p\,\bar d_p-\bar d_p\,d^p)\,\Big\}-\cr
&&~~~~~~~~~~~~~-~  {\frac {1}{2}}\,{\frac {\bar\partial_{pq}}{\parp}}\,{\biggl \{}\,\frac{\parp}{{\sqrt 2}}\,{{ \theta}^p}\,{{ \theta}^q}\,-\,\,\frac{\sqrt 2}{\parp}\,{{\partial}^p}\,{{ \partial}^q}+\frac{1}{\sqrt{2}\parp}d^p\,d^q\,\,{\biggr \}}\ ,
\eea
and its conjugate~\cite{ABR1}. These yield the linear parts of the dynamical supersymmetry generators.
\bea
\label{dsg}
{\cal Q}^m_{}&=&{\frac {\bar \partial}{\parp}}\,{{q_+}^m}\,+\,{\frac {\partial^{mn}}{\parp}}\,{{\bar q}_{+\,n}}\ ,\cr
&&\cr
\bar{\cal Q}_m^{}&=&\frac {\partial}{\parp}}\,{{\bar q}_{+\,m}}\,+ \,{\frac {{\bar \partial}_{mn}}{\parp}\, q^{~n}_{\,+}\ ,
\eea
which satisfy
\be
\{\,{\cal Q}^{\,m}_{}\,,\,\bar {\cal Q}^{}_{\,n}\,\}~=~i\,\sqrt{2}\,\frac{1}{\parp}
\,\Big(\delta^m_{~n}\,\partial\,{\bar \partial}\,+\,{\bar \partial}_{mp}\,\partial^{np}\,\Big)\ ,
\ee
\vskip 0.3cm
\ndt Third, the introduction of a `generalized' derivative
\be
{\bar \nabla}~\equiv~{\bar \partial}\,+\,\frac{i}{4\,\sqrt 2\,\partial^+}\,{{\bar d}_p}\,{{\bar d}_q}\,\partial^{pq} \ ,
\ee
and its conjugate. 
\vskip 0.3cm
\ndt The kinetic term is trivially made $SO(8)$-invariant by including the six extra transverse derivatives in the d'Alembertian. The quartic interactions are  obviously invariant since they  do not contain any transverse derivative operators. Hence we need only consider the cubic vertex. 
\vskip 0.3cm
\ndt The essence of~\cite{ABR1} was that covariance in ten dimensions is achieved by simply replacing transverse derivatives $\partial$ and $\bar\partial$ by $\nabla$ and $\bar\nabla$, respectively. This leads to the following cubic interaction term in ten dimensions~\cite{ABR1} 
\be
\label{tend}
\int d^{10}x\int d^4\theta\,d^4 \bar \theta\,{\cal L}_{10}\ ,
\ee
where
\bea
{\cal L}_{10}&=&-\bar\phi\,\frac{\Box_{10}}{\partial^{+2}}\,\phi
~+\frac{4g}{3}\,f^{abc}_{}\,\Big(\frac{1}{\partial^+_{}}\,\bar\phi^a_{}\,\phi^b_{}\,\bar\nabla\,\phi^c_{}+{\rm complex~conjugate}\Big)\cr
&&-g^2f^{abc}_{}\,f^{ade}_{}\Big(\,\frac{1}{\partial^+_{}}(\phi^b\,\partial^+\phi^c)\frac{1}{\partial^+_{}}\,(\bar \phi^d_{}\,\partial^+_{}\,\bar\phi^e)+\frac{1}{2}\,\phi^b_{}\bar\phi^c\,\phi^d_{}\,\bar\phi^e\Big)\ .
\eea

\section{New results}

We have seen that the ($\mathcal N=4,d=4$) Hamiltonian is a quadratic form \eqref{4hamil2}. In the following, we will prove that the Hamiltonian for the ten-dimensional theory described by (\ref {tend}) is also a quadratic form.

\subsection{The kinetic term} 
\label{ssec : qfkinetic}

\ndt  Starting from the free dynamical supersymmetry generators in (\ref {dsg}) and adding the non-linear term in  (18) we find the full non-linear dynamical supersymmetry generators to be.
\bea
\delta_{q_-{}^m}\bar \phi^a=Q^m\,\bar\phi^a-gf^{abc}\fr{\partial^+}(q_+^m\bar\phi^b\partial^+\bar\phi^c)\ , \nn \\
\delta_{\bar q_-{}_m} \phi^a={\bar Q}_m\,\phi^a-gf^{abc}\fr{\partial^+}({\bar q}_{+m}\,\phi^b\partial^+\phi^c)\ ,
\eea
where we remind the reader that fields within this superfield now depend on all ten coordinates.
\vskip 0.3cm
\ndt Our claim is that the ten-dimensional Hamiltonian of $\mathcal N=1$ Yang-Mills is simply
\be
\label{10hamil2}
H\,=\,\frac{i}{\sqrt 2}\int{d^9}x\,{d^4}\theta\,{d^4}{\bar \theta}\, \delta_{q^m}\bar \phi^a \,\frac{1}{\partial^+} \,\delta_{\bar q_m} \phi^a.
\ee

\ndt We start by verifying this claim, at the free level
\begin{eqnarray}
 \delta_{q^m}\bar \phi^a \,\frac{1}{\partial^+} \,\delta_{\bar q_m} \phi^a&=&\{(\frac{\bar{\partial}}{\partial^+}q^m_+\, +\,\frac{\partial^{mn}}{\partial^+}\, \bar{q}_{+n})\,\bar{\phi}^a\, \frac{1}{\partial^+}(\frac{{\partial}}{\partial^+}\bar{q}_{+m}\, +\,\frac{\bar{\partial}_{mp}}{\partial^+} {q}^p_+)\, \phi^a \} \nonumber \\
&=& \{\frac{\bar{\partial}}{\partial^+}q^m_+\,\bar{\phi}^a\,\frac{{\partial}}{{\partial^+}^2}\bar{q}_{+m}\,\phi^a \,+ \,\frac{\partial^{mn}}{\partial^+} \bar{q}_{+n}\bar{\phi}^a\frac{{\partial}}{{\partial^+}^2}\bar{q}_{+m}\phi^a \nonumber \\
&&\,\,\,+\, \frac{\bar{\partial}}{\partial^+}q^m_+\,\bar{\phi}^a\,\frac{\bar{\partial}_{mp}}{{\partial^+}^2} {q}^p_+ \,\phi^a \,+\,\frac{\partial^{mn}}{\partial^+} \bar{q}_{+n}\,\bar{\phi}^a\,\frac{\bar{\partial}_{mp}}{{\partial^+}^2} {q}^p_+ \,\phi^a \}\nonumber \\
&=& \{\mathcal{A}+\mathcal{B}+\mathcal{C}+\mathcal{D} \}  \
\end{eqnarray}
We focus first on term $\mathcal B$ which after integration by parts and use of the `inside-out' constraint in (\ref{ioc}) yields
\begin{equation}
\mathcal{B}\,=\,-\,\frac{1}{2}\,\bar{\phi}^a\,\frac{\partial^{mn}\partial}{{\partial^+}^3}\,\{\bar{q}_{+n},\bar{q}_{+m}\}\,\phi^a\,=\,0\
\end{equation}
The term $\mathcal C$ vanishes in an identical manner. The non-vanishing contributions come from terms $\mathcal A$ and $\mathcal D$. Term $\mathcal A$ after similar simplification becomes
\begin{equation}
\mathcal{A}\,=\,-\,\frac{1}{2}~\bar{\phi}^a~\frac{\partial\bar{\partial}}{{\partial^+}^3} ~\{q^m_+,\bar{q}_{+m}\}~\phi^a~=~-i2~\sqrt{2}\,\bar{\phi}^a~\frac{\partial\bar{\partial}}{{\partial^+}^2}~\phi^a\ ,
\end{equation}
while $\mathcal{D}$ reads
\begin{equation}
\mathcal{D}~=~-\,\frac{i}{\sqrt{2}}~\bar{\phi}^a~\frac{\partial^{mn}\bar{\partial}_{mn}}{{\partial^+}^2}~\phi^a
\end{equation}
Thus the free ten-dimensional Hamiltonian reads
\begin{equation}
H~=~2~\bar{\phi}^a~\frac{\partial\bar{\partial}}{{\partial^+}^2}~\phi^a+\frac{1}{2}~\bar{\phi}^a~\frac{\partial^{mn}\bar{\partial}_{mn}}{{\partial^+}^2}~ \phi^a\ ,
\end{equation}
as expected, given the Lagrangian in (\ref {tend}).

\subsection{The cubic interaction vertex}

Having shown that the free Hamiltonian is a quadratic form, we now move to examining the cubic interaction vertex. The relevant piece from (\ref {10hamil2}) is
\bea
 \delta_{q^m}\bar \phi^a \,\frac{1}{\partial^+} \,\delta_{\bar q_m} \phi^a|_g=-\,f^{abc}\fr{\partial^+}(q_+^m\bar\phi^b\partial^+\bar\phi^c)\,\fr{\partial^+}{\bar Q}_m\,\phi^a\ .
\eea
\ndt We only need to focus on the term that involves the new transverse derivatives. This may be written as 
\begin{eqnarray} \label{eq:2.1}
&&f^{abc} \int \frac{\bar{\partial}_{mn}}{\partial^{+2}} q^n_+\phi^a \frac{1}{\partial^+}(q^m_+\bar{\phi}^b\partial^+\bar{\phi}^c) \nonumber  \\
&&=\frac{1}{3} f^{abc}\int \left( \frac{1}{\partial^+}\phi^a \bar{\phi}^b \frac{d^m d^n \bar{\partial}_{mn}}{\partial^+}\bar{\phi}^c \, -\, \frac{1}{2}\phi^a \bar{\phi}^b \frac{d^m d^n \bar{\partial}_{mn}}{\partial^{+2}}\bar{\phi}^c  \right) \ .
\end{eqnarray}
\ndt We briefly review the derivation of the above result. We start with the explicit form of $q_+$ and write the L.H.S. of \eqref{eq:2.1} as 
\begin{equation}\label{eq:2.2}
-i\sqrt{2} f^{abc} \int \frac{\bar{\partial}_{mn}}{\partial^{+2}}\phi^a\theta^m\partial^n\bar{\phi}^b\partial^+\bar{\phi}^c \
\end{equation}
Partial integration with respect to $\partial^n$ ($f^{abc}$ and the integral sign are suppressed) gives
\begin{equation}\label{eq:2.3}
i\sqrt{2}\,\theta^m\partial^n \frac{\bar{\partial}_{mn}}{\partial^{+2}}\phi^a\bar{\phi}^b\partial^+\bar{\phi}^c+ i\sqrt{2} \, \frac{\bar{\partial}_{mn}}{\partial^{+2}} \phi^a \bar{\phi}^b\partial^+\theta^m\partial^n\bar{\phi}^c \equiv \rom{1} +\rom{2}
\end{equation}
Using the `inside out' constraint and partially integrating the chiral derivatives, the first term of \eqref{eq:2.3} is
\begin{eqnarray}\label{eq:2.4}
\rom{1}&=&\frac{i\sqrt{2}}{2\cdot4!} (\epsilon^{ijkl}\bar{d}_i\bar{d}_j\bar{d}_k\bar{d}_l) \theta^m\partial^n\frac{\bar{\partial}_{mn}}{\partial^{+2}}\phi^a\bar{\phi}^b\frac{1}{\partial^+}\phi^c \nonumber\\
&=&-\frac{1}{2}\frac{d^m d^n\bar{\partial}_{mn}}{\partial^+} \bar{\phi}^a \bar{\phi}^b \frac{1}{\partial^+}\phi^c + i\sqrt{2}\, \theta^m \partial^n \bar{\partial}_{mn}\bar{\phi}^a \bar{\phi}^b\frac{1}{\partial^+}\phi^c\
\end{eqnarray}
Similar manipulation of the second term in \eqref{eq:2.3} yields
\begin{equation}\label{eq:2.5}
\rom{2}=-i\sqrt{2} \, \frac{\bar{\partial}_{mn}}{\partial^+} \phi^a \bar{\phi}^b \theta^m\partial^n\bar{\phi}^c - i\sqrt{2} \, \frac{\bar{\partial}_{mn}}{\partial^{+2}} \phi^a \partial^+\bar{\phi}^b \theta^m\partial^n\bar{\phi}^c \
\end{equation}
Integration by parts  with respect to $\bar{\partial}_{mn}$ in the first term of \eqref{eq:2.5} gives 
\begin{equation} \label{eq:2.6}
i\sqrt{2}\, \frac{1}{\partial^+} \phi^a \bar{\partial}^{mn}\bar{\phi}^b \theta^m \partial^n \bar{\phi}^c + i \sqrt{2}\, \frac{1}{\partial^+}\phi^a\bar{\phi}^b\bar{\partial}_{mn}\theta^m \partial^n \bar{\phi}^c\ ,
\end{equation}
with the second term cancelling against the second term of \eqref{eq:2.4}. Using `inside out' constraints on $\bar{\phi}^b$, the first term of \eqref{eq:2.6} becomes 
\begin{equation} \label{eq:2.7}
i \sqrt{2}\, \partial^+\bar{\phi}^a \frac{\bar{\partial}_{mn}}{\partial^{+2}}\phi^b \theta^m \partial^n \bar{\phi}^c +\frac{1}{2} \phi^a \bar{\phi}^c \frac{\bar{\partial}_{mn}}{\partial^{+2}}d^m d^n \bar{\phi}^b - \frac{1}{2}\frac{1}{\partial^+} \phi^a \bar{\phi}^c \frac{\bar{\partial}_{mn}}{\partial^+}d^m d^n \bar{\phi}^b
\end{equation}
\ndt So 
\begin{eqnarray} \label{eq:2.8}
\rom{1}+\rom{2}&=&i2 \sqrt{2}\, \frac{\bar{\partial}_{mn}}{\partial^{+2}}\phi^a\theta^m\partial^n\bar{\phi}^b\partial^+\bar{\phi}^c \nonumber \\
&&+\frac{1}{\partial^+}\phi^a \bar{\phi}^b \frac{d^m d^n \bar{\partial}_{mn}}{\partial^+}\bar{\phi}^c \, -\, \frac{1}{2}\phi^a \bar{\phi}^b \frac{d^m d^n \bar{\partial}_{mn}}{\partial^{+2}}\bar{\phi}^c 
\end{eqnarray}
\vskip 0.3cm
\ndt The R.H.S. above is clearly equal to \eqref{eq:2.2} (from which we started) and this leads to \eqref{eq:2.1}.
\vskip 0.3cm
\ndt Following the same procedure, we obtain the conjugate of \eqref{eq:2.1} 
\begin{eqnarray}\label{eq:2.9} 
&&f^{abc}\int \frac{\partial^{mn}}{\partial^+}\bar{q}_{+n}\bar{\phi}^a\frac{1}{\partial^{+2}}(\bar{q}_{+m}\phi^b\partial^+\phi^c) \nonumber \\
&&=\frac{1}{3} f^{abc} \int \left(\frac{1}{\partial^+}\bar{\phi}^a\phi^b\frac{\bar{d}_m\bar{d}_n\partial^{mn}}{\partial^+}\phi^c-\frac{1}{2}\bar{\phi}^a \phi^b \frac{\bar{d}_m\bar{d}_n\partial^{mn}}{\partial^{+2}}\phi^c\right)\ .
\end{eqnarray} 
Using the inside-out constraint on $\phi^c$ in the second term of \eqref{eq:2.9} we find
\begin{equation}
f^{abc}\bar{\phi}^a \phi^b \frac{\bar{d}_m\bar{d}_n\partial^{mn}}{\partial^{+2}}\phi^c=-\, f^{abc}\phi^a \bar{\phi}^b \frac{d^m d^n \bar{\partial}_{mn}}{\partial^{+2}}\bar{\phi}^c\ .
\end{equation}
Thus, the sum of \eqref{eq:2.1} and \eqref{eq:2.9} is 
\begin{eqnarray}
&&f^{abc} \int \left\{\frac{\bar{\partial}_{mn}}{\partial^{+2}} q^n_+\phi^a \frac{1}{\partial^+}(q^m_+\bar{\phi}^b\partial^+\bar{\phi}^c)+ \frac{\partial^{mn}}{\partial^+}\bar{q}_{+n}\bar{\phi}^a\frac{1}{\partial^{+2}}(\bar{q}_{+m}\phi^b\partial^+\phi^c) \right\} \nonumber \\
&&= \frac{1}{3} f^{abc}\int \left( \frac{1}{\partial^+}\phi^a \bar{\phi}^b \frac{d^m d^n \bar{\partial}_{mn}}{\partial^+}\bar{\phi}^c \,+\, \frac{1}{\partial^+}\bar{\phi}^a\phi^b\frac{\bar{d}_m\bar{d}_n\partial^{mn}}{\partial^+}\phi^c \right) \ ,
\end{eqnarray}
which is an exact match to what is expected from (\ref {tend}). The repeated use of the inside-out constraints in all these computations clearly suggests that maximal supersymmetry is essential to many of the simplifications presented here.
\vskip 0.3cm

\subsection{The quartic interaction vertex}

\ndt We do not need to check the quartic interaction vertex since it does not involve any transverse derivatives. This means that the results in~\cite{ABKR} for the quartic vertex carry over to our case with the two standard modifications used in this section: the fields now depend on all ten directions and the space-time integration is over all ten coordinates.
\vskip 0.5cm

\section{Pure Yang-Mills}

Having established that both the ($\mathcal N=4,d=4$) and ($\mathcal N=1, d=10)$ theories may be written as quadratic forms, we now turn to the case of pure Yang-Mills theory. We ask the same question in this case, whether the Hamiltonians describing the interacting theories can be written as quadratic forms.

\subsection{d=4}
\ndt The Lagrangian for pure Yang-Mills theory may be read of from our earlier results~(\ref {fourdd})
\bea 
\label{fourdh}
\mathcal L \,&=&\, \bar{A}^a\,\Box\, A^a\,-\, 2\,g\, f^{abc}\,(\frac{\bar{\partial}}{\partial^+}\,A^a\partial^+\bar{A}^b\,A^c\,+\,\frac{\partial}{\partial^+}\,\bar{A}^a\,\partial^+\,A^b\,\bar{A}^c) \nn\\
&&-\, 2\,g^2\,f^{abc}\,f^{ade}\,\frac{1}{\partial^+}(\partial^+\,A^b\,\bar{A}^c)\,\frac{1}{\partial^+}(\partial^+\,\bar{A}^d\,A^e)\ ,
\eea
with the corresponding Hamiltonian
\bea \label{eq:HamilYM}
\mathcal H \,&=&\, \bar{A}^a\,\bar{\partial}\partial\, A^a\,+\,g\, f^{abc}\,(\frac{\bar{\partial}}{\partial^+}\,A^a\partial^+\bar{A}^b\,A^c\,+\,\frac{\partial}{\partial^+}\,\bar{A}^a\,\partial^+\,A^b\,\bar{A}^c) \nn\\
&&-\, g^2\,f^{abc}\,f^{ade}\,\frac{1}{\partial^+}(\partial^+\,A^b\,\bar{A}^c)\,\frac{1}{\partial^+}(\partial^+\,\bar{A}^d\,A^e)\ .
\eea
We introduce the following derivative structure
\be
\label{dstruct}
\mathcal{\bar{D}}\,A^a\,\equiv \,\bar{\partial}A^a\,-\,g\,f^{abc}\,\frac{1}{\partial^+}(\bar{A}^b\partial^+\,A^c)\ ,
\ee
which allows us to recast the Hamiltonian as
\be \label{Yhamil4}
\mathcal H\,=\,-\, \int d^3x \mathcal D\bar{A}^a\,\mathcal{\bar{D}}A^a\ ,
\ee
yielding a quadratic form.

\vskip 0.3cm

\subsection{A note on gauge invariance}

We know that when we fix a gauge in Yang-Mills theory, there is residual gauge invariance. One could then ask if the form of \eqref{Yhamil4} is governed by some such residual gauge symmetry. This fact was discussed recently~\cite{ABK} for gravity but the situation is similar here. Choosing the gauge $A^+=0$ implies some remaining gauge invariance, with a gauge parameter that satisfies $\partial^+ \Lambda = 0$. However when we solve for the unphysical degree of freedom $A^-$ we fix an integration parameter and it is often said that the gauge has been completely chosen. It is true that there is no infinitesimal symmetry that can be integrated to a finite symmetry. There is however still an infinitesimal one satisfying  $\partial^+ \Lambda = 0$ as well as $\bar \partial \partial \Lambda = 0$. This is sufficient to determine that \eqref{Yhamil4} is of the form
\be \label{prop1}
\mathcal{\bar{D}}\,A^a\,\equiv \,\bar{\partial}A^a\,-\,g\,f^{abc}\,\frac{1}{\partial^+{}^n}(\bar{A}^b\partial^+{}^n\,A^c)\ .
\ee
To determine the value $n=1$, we need to check Poincar\'e invariance. In fact we can now truly regard the operator $\mathcal{\bar{D}}$ as a covariant derivative and we find that the expression for the Hamiltonian is indeed invariant under the remaining gauge invariance.
\vskip 0.3cm
\ndt One might ask how the expression (\ref{eq:nlinear}) can be covariant under the remaining gauge invariance when the superfield $\phi^a$ involves both $A^a$ and $\bar A^a$. The answer is that the superfield $\delta^{}_{\bar q_+{}_m}\,\phi^a$ cleverly only involves $\bar A^a$ and expression (\ref{eq:nlinear}) may be regarded as the covariant derivative of the superfield.

\vskip 0.3cm

\subsubsection{Non-helicity basis}
All our work in this paper, thus far, has been in a helicity basis. We could however, have chosen to work in a non-helicity basis as was done in~\cite{SS}. To reach their Lagrangian from (\ref {fourdh}) we simply use
\bea
A^a\,&=&\,\frac{1}{\sqrt{2}}(A_1^a\,+i\,A_2^a)\ ;\qquad \, \bar{A}^a\,=\,\frac{1}{\sqrt{2}}(A_1^a\,-i\,A_2^a)\ ,
\eea
and the Lagrangian is then
\bea
\label{eq:FlagYM}
\mathcal L& =& \frac{1}{2} A^a_i \Box A^a_i\,-\, 2\,g\, f^{abc}\,\frac{1}{\partial^+}A^a_i\,\partial_i\,A^b_j\,\partial^+\, A^c_j\,-\,\frac{1}{4}\,g^2\,f^{abc}\,f^{ade}\, A^b_i\,A^c_j\,A^d_i\,A^e_j \nonumber\\
&&-\,\frac{1}{2}\, g^2\, f^{abc}\, f^{ade}\,\frac{1}{\partial^+}(A^b_i\,\partial^+\,A^c_i)\,\frac{1}{\partial^+}(A^d_j\,\partial^+\,A^e_j)\ .
\eea
\ndt There are obviously two gauge covariant forms that we can write under the remaining gauge invariance. One is the expression \eqref{prop1} and the other is 
\be
F_{ij}{}^a = \partial_i A_j{}^a - \partial_j A_i{}^a - g f^{abc} \frac{1}{{\partial^+}^n}(A_i{}^b {\partial^+}^n A_j{}^c)+g f^{abc} \frac{1}{{\partial^+}^n}(A_j{}^b {\partial^+}^n A_i{}^c)\ .
\ee
\ndt We compare them and find
\be
\mathcal{\bar{D}}\,A^a = \frac{1}{2} (\partial_i A_i{}^a - g f^{abc}\frac{1}{\partial^+}(A_i{}^b\partial^+A_i{}^c) )+\frac{i}{4} \epsilon^{ij}F_{ij}{}^a\ .
\ee
It is obvious that the Hamiltonian cannot be written in terms of $ F_{ij}^a$. Hence it is only in a helicity base that the Hamiltonian can be written as a quadratic form.
\vskip 0.3cm

\vskip 0.5cm

\subsection{d=10}

\ndt There is no helicity basis in ten dimensions. The Lagrangian (and hence the Hamiltonian) for $d=10$ Yang-Mills has the same structure as the $ d=4$ case in~\eqref{eq:FlagYM} (with $i,j=1...8$).  By the same arguments presented here, the pure Yang-Mills Hamiltonian in $d=10$ cannot be expressed as a quadratic form.

\vskip 0.5cm

\section{Conclusions}

\ndt We have found that the Hamiltonian of $\mathcal N=4$ Yang-Mills theory in the light-cone gauge formulation, its 'oxidized' parent theory in $d=10$ as well as the $d=4$ pure Yang-Mills theory all take particularly simple forms when formulated in terms of helicity. This is not totally unexpected, given that amplitude analyses in terms of helicity have proven extremely useful~\cite{Parke:1986gb, Witten:2003nn}. Here we see evidence of this at a fundamental level. 
\vskip 0.3cm
\ndt Can these facts teach us more about these field theories? The form can be used in the algebraic formulation where the generators of the non-linear super Poincar\'e algebra are constructed \cite{Bengtsson:1983pg}. However, these are well-known and worked through in the original papers and nothing new will be learnt here. The original formulation \cite{Brink:1982pd} was also essential in the proof of perturbative finiteness of the $\mathcal N=4$ theory~\cite{Brink:1982wv}. We do not see any advantage here of using the new formulation since the proof was performed by studying detailed properties of general Feynman diagrams. There should be an advantage though to using this formalism in more general studies of scattering amplitudes but we have not done so yet.
\vskip 0.3cm
\ndt Could this formalism tell us about further unknown symmetries? In the  light-cone formulation of the  $\mathcal N=8$ supergravity theory, the quadratic form was important in understanding the $E_{7(7)}$ invariance as a non-linear $\sigma$-type symmetry \cite{Brink:2008qc}. For the $\mathcal N=4$ case, with a dimensionless coupling constant there is no possiblilty of a $\sigma$-type symmetry. Instead it highlights the importance of residual gauge invariance. For vector particle scattering, there must be derivatives acting on the external lines making the amplitudes behave even more convergently~\cite{Kallosh:2010mk} than predicted by pure power counting. Because of supersymmetry that must also be the case for all scattering amplitudes. This situation is analogous to Delbr\"uck scattering.
\vskip 0.3cm
\ndt We have completed our study of quadratic forms in both the supersymmetric and pure Yang-Mills systems. It would be interesting to see what happens to these quadratic forms when we attempt a systematic truncation of supersymmetry in these theories~\cite{BT}. Perhaps similar quadratic form structures also appear in four-dimensional higher spin theories in light-cone gauge~\cite{HST}. As stated in the introduction, one major follow-up would be to carry  over this study to the gravity-supergravity system and see what this can teach us about the symmetries that abound there and $\mathcal N=8$ supergravity in particular.

\vskip 0.5cm
\ndt {\it {Acknowledgments}}
\vskip 0.3cm

\ndt We thank Marc Henneaux, Hermann Nicolai, Sarthak Parikh and Pierre Ramond for helpful discussions. 

\end{document}